\documentstyle[12pt]{ioplppt}                  
\jl{6}                                         

\begin{document}

\def \D {\mbox{D}}
\def \d {\mbox{d}}
\def \div {\mbox{div}\,}
\def \c {\mbox{curl}\,}
\def \ep {\varepsilon}
\def \tr {\mbox{tr}\,}
\def \ts {\textstyle}
\def \rd {\displaystyle{\cdot}}

\title{Newtonian-like and anti-Newtonian universes}

\author{
Roy Maartens\dag\,,
William M Lesame\ddag\ and
George F R Ellis\S
}

\address{\dag\ School of Computer Science and Mathematics, 
Portsmouth University, Portsmouth PO1~2EG, Britain}

\address{\ddag\ Department of Mathematics, University of South Africa,
Pretoria~0001, South Africa}

\address{\S\ Department of Mathematics and Applied Mathematics,
University of Cape Town, Cape Town~7700, South Africa}

\begin{abstract}

In an irrotational dust universe, the locally free gravitational
field is covariantly described
by the gravito-electric and gravito-magnetic tensors $E_{ab}$ and
$H_{ab}$. In Newtonian theory, $H_{ab}=0$
and $E_{ab}$ is the tidal tensor. Newtonian-like dust universes in
general relativity (i.e. with $H_{ab}=0$, often called `silent')
have been shown to be inconsistent in general and unlikely to
extend beyond the known spatially homogeneous or Szekeres examples.
Furthermore, they are subject to 
a linearization instability. Here we show that `anti-Newtonian'
universes, i.e. with purely gravito-magnetic field, so that
$E_{ab} = 0\neq H_{ab}$, are also subject to severe integrability 
conditions. Thus these models are inconsistent in general. 
We show also that there are no anti-Newtonian spacetimes 
that are linearized perturbations of Robertson-Walker universes.
The only $E_{ab}=0\neq H_{ab}$ solution known to us is not a dust
solution, and we show that it is kinematically G\"{o}del-like but
dynamically unphysical.

\end{abstract}

\pacs{0420, 9880, 9530}


\section{Introduction}

Irrotational dust spacetimes are characterized by vanishing
pressure ($p=0$)\footnote{It is implicit that the anisotropic 
stress $\pi_{ab}$ and the energy flux $q_a$ also vanish.}
and vorticity ($\omega_a=0$), and positive
energy density ($\rho > 0$). They
are important arenas for studying both
the late universe \cite{enm,mm}
and gravitational collapse models \cite{bj}. 
Apart from the spatially homogeneous and isotropic 
Friedmann-Lemaitre-Robertson-Walker (FLRW) case, 
which is characterized by vanishing shear ($\sigma_{ab}=0$),
these spacetimes
have non-zero shear and non-zero locally
free gravitational field. This field
is represented by the irreducible electric and magnetic parts
$E_{ab}$ and $H_{ab}$ of the Weyl tensor (see \cite{mes} for 
a covariant analysis of local freedom in the gravitational
field).

The gravito-electric tensor $E_{ab}$ is the relativistic 
generalization of the tidal tensor in Newtonian theory, while 
the gravito-magnetic tensor $H_{ab}$ has no
Newtonian analogue \cite{e,ed}, and is associated with gravitational
radiation. There are no sound waves in dust, and
when $H_{ab}=0$ there can be no gravitational radiation \cite{he,dbe}.
Dust spacetimes with $H_{ab}=0$ have therefore been called
`silent', since there are no propagating
signals \cite{mps,cpss,bmp}.\footnote{A weaker covariant condition
for the absence of gravitational radiation is that the spatial
curls (defined below) of $E_{ab}$ and $H_{ab}$ must vanish 
\cite{he,dbe,mes}.}
Within the class of irrotational dust models, the
silent ($H_{ab}=0$) universes may be called {\em `Newtonian-like'\,}, 
since they have a clear
Newtonian counterpart. One might therefore
expect that there is a broad variety of such Newtonian-like 
spacetimes,
representing the general relativistic generalization of simple
Newtonian models.
However, there are subtleties involved in
the Newtonian limit of general relativity (see for example
\cite{ed,e2,kp,sss}), and it turns out
that the relativistic Newtonian-like 
spacetimes form a very restricted class.

Independent exact analyses 
of silent $H_{ab}=0$ models, using different methods, are
given in \cite{5} and \cite{s}. The independent approaches 
produce the same conclusion. The silent
condition $H_{ab}=0$ leads to an integrability condition whose
repeated time differentiation forms in general
a non-terminating chain 
that leads to inconsistencies. It is thus unlikely that 
consistent silent solutions exist beyond 
the known special cases of Szekeres spacetimes
and spatially homogeneous models \cite{5,s}. Furthermore,
since these integrability conditions are identically satisfied 
in the linearized theory (with FLRW background), 
{\em Newtonian-like models have
a linearization instability\,} 
\cite{mn,5}, i.e. there are consistent
linearized solutions which are not the limit of any consistent
solutions in the full, nonlinear theory.

The linearization
instability found via an exact covariant analysis in
\cite{mn,5} is confirmed
by an independent analysis based on a $1/c$ expansion
of the field equations and Bianchi identities \cite{kp}. 
The gravito-magnetic tensor vanishes in the Newtonian, but not the
post-Newtonian, limit. Forcing $H_{ab}=0$ at the
post-Newtonian level has the consequence that
the non-local nature of the Newtonian tidal force cannot be
recovered in the Newtonian limit, and reflects the
fact that it is incorrect in general to assume $H_{ab}=0$
in general relativity.

Thus the Newtonian-like silent models ($H_{ab}=0$)
have a very narrow and limited applicability in cosmology and 
especially in gravitational instability \cite{kp,hb,5}. 
Realistic
collapse scenarios or realistic inhomogeneous models of the late
universe require a gravito-magnetic field $H_{ab}$.
Spacetimes with $H_{ab}\neq0$ do not have a Newtonian 
counterpart \cite{ed},
and they are consistent in the generic case, i.e. when
there is also a gravito-electric field $E_{ab}$ and 
no external conditions are enforced on $H_{ab}$ or $E_{ab}$ \cite{m}.

The most `extreme' non-Newtonian models (in terms of `distance' from 
Newtonian theory) are those with purely gravito-magnetic
field, i.e. with $E_{ab}=0\neq H_{ab}$. (Note that $E_{ab}=0$ and
$H_{ab}=0$ together
imply that the spacetime is FLRW.) Consequently, we will use the term
{\em `anti-Newtonian'\,} for irrotational dust spacetimes with
$E_{ab}=0\neq H_{ab}$.
Purely gravito-magnetic 
spacetimes (not restricted to irrotational dust)
appear to have first been discussed in \cite{t},
where it was shown that either the shear or the vorticity must be
non-vanishing.
In this paper we consider only such models 
containing irrotational dust. Since $E_{ab}=0$ also implies no
gravitational radiation \cite{he}, the anti-Newtonian models are
also `silent'.

Here we show that irrotational dust 
spacetimes with $E_{ab}=0$
are subject to integrability conditions that are even more
restrictive than in the $H_{ab}=0$ case. The
integrability conditions once again form in general
non-terminating chains that
lead to inconsistencies. There may be spatially
homogeneous models of this type which satisfy the 
integrability conditions,
but we have been unable to find examples. However, a further result
indicates that there are unlikely to be any consistent 
exact solutions. This further result is that the only linearized 
irrotational dust solutions with $E_{ab}=0$ are exactly FLRW 
(which have $H_{ab}=0$) -- i.e. {\em there are no linearized 
anti-Newtonian irrotational dust models.}

The question arises as to whether there are any known purely
gravito-magnetic solutions ($E_{ab}=0\neq H_{ab}$)
at all. In \cite{mawh}, it is conjectured
that there are no such non-flat vacuum solutions.
An exact non-vacuum solution is given in \cite{almp},
apparently the first purely gravito-magnetic solution. 
However, as pointed out in \cite{almp},
this solution has unphysical Segre type.
We show in an appendix 
that it is kinematically a magnetic counterpart of
the G\"odel solution (which is purely gravito-electric \cite{e}), but
that dynamically it has an unphysical source, with negative
energy density.

In section 2 we discuss briefly the covariant propagation and 
constraint equations, and the covariant approach that has been
developed to analyze consistency \cite{m}.
We show in section 3 how the silent condition $H_{ab}=0$ implies
a primary covariant integrability condition, which is the basis for
the linearization instability, as well as for the indefinite chain of 
conditions in the nonlinear case. 
This discussion forms a prelude to the
analysis in section 4 of the $E_{ab}=0$ case, where there are two
primary integrability conditions, each of which produces in
general an indefinite chain
of further conditions after differentiation. In the linearized 
theory, the primary conditions themselves lead to the vanishing of 
anisotropy and inhomogeneity, i.e. to the FLRW case, so that
there are no linearized anti-Newtonian models.

We use the notation and conventions of \cite{m,mes}. Units
are such that
$8\pi G=1=c$; $a,b,\cdots$ are spacetime indices;
(square) round
brackets enclosing indices denote (anti-)
symmetrization, while
angled brackets denote the spatially projected, symmetric and 
tracefree part.

\section{Covariant dynamical equations}

A covariant approach to the propagation and constraint equations
and their consistency has been developed in \cite{m}.
There it was shown that in generic irrotational
dust spacetimes, i.e. without restrictions on $E_{ab}$ and $H_{ab}$,
the constraints are consistent and evolve 
consistently with the propagation equations. 
(See also \cite{mle,l}.)
However, as discussed above and shown below, 
consistency breaks down in general when the
Newtonian condition $H_{ab}=0$ or the anti-Newtonian condition
$E_{ab}=0\neq H_{ab}$ are imposed.

The dust four-velocity $u^a$ 
(with $u^au_a=-1$) provides a unique covariant $1+3$
splitting, as fully discussed in \cite{e}. Here we follow the
streamlined version of the formalism developed in \cite{m,mes}.
The Weyl tensor splits covariantly into gravito-electric
and gravito-magnetic parts\footnote{See \cite{mb} for a thorough
discussion of gravito-electromagnetism.}
\[
E_{ab}=C_{acbd}u^c u^d\,,~~ H_{ab}={\ts{1\over2}}\ep_{acd}
C^{cd}{}{}_{be}u^e\,,
\]
where the spatial
permutation tensor is
$\ep_{abc}=\eta_{abcd}u^d$, and $\eta_{abcd}$ is the spacetime
permutation tensor. 
The tensor $h_{ab}=g_{ab}
+u_au_b$, where $g_{ab}$ is the metric tensor, projects orthogonal to
$u^a$. Then the projected tracefree symmetric part of a rank-2
tensor is
$S_{\langle ab\rangle }=h_a{}^c h_b{}^d S_{(cd)}-
{\ts{1\over3}}S_{cd}h^{cd} h_{ab}$. The fluid kinematics and dynamics, 
and the locally free field are described by the covariant scalars
$\rho$ (energy density) and $\Theta$ (volume expansion rate), 
and rank-2 
tensors $\sigma_{ab}=\sigma_{\langle ab\rangle}$ (shear),
$E_{ab}=E_{\langle ab\rangle}$ and $H_{ab}=
H_{\langle ab\rangle}$. 
The pressure $p$ and the vorticity                                 
$\omega_a$ are assumed to vanish in the irrotational
dust case we consider here.                 

The covariant derivative splits into a
covariant time derivative 
$\,\dot{S}_{a\cdots}=u^b\nabla_b S_{a\cdots}\,$,
and a covariant spatial derivative 
$\,\D_aS_{b\cdots}=h_a^ch_b^d \cdots\nabla_cS_{d\cdots}\,$.
Then the latter leads to a covariant spatial divergence and 
curl \cite{m}:
\begin{eqnarray*}
\div V=\D^aV_a\,,~ &&~ \c V_a=\ep_{abc}\D^bV^c \,, \\
(\div S)_a=\D^b S_{ab} \,, ~ && ~
\c S_{ab}=\ep_{cd(a}\D^c S_{b)}{}^d \,.
\end{eqnarray*}
Important identities obeyed by these derivatives are collected in
Appendix A.

Covariant splitting of the Bianchi identities and the Ricci identity
for $u^a$, where the field equations are taken as an algebraic
definition of the Ricci tensor, leads to the
propagation equations
\begin{eqnarray}
\dot{\rho}+\Theta\rho &=& 0 \,,
\label{p1}\\
\dot{\Theta}+{\ts{1\over3}}\Theta^2 &=& -{\ts{1\over2}}\rho
-\sigma_{ab}\sigma^{ab} \,,
\label{p2}\\
\dot{\sigma}_{ab}+{\ts{2\over3}}\Theta\sigma_{ab}+\sigma_{c\langle a}
\sigma_{b\rangle }{}^c &=& -E_{ab} \,,
\label{p3}\\
\dot{E}_{ab}+\Theta E_{ab}-3\sigma_{c\langle a}E_{b\rangle }{}^c &=&
\c H_{ab}-{\ts{1\over2}}\rho\sigma_{ab} \,,
\label{p4}\\
\dot{H}_{ab}+\Theta H_{ab}-3\sigma_{c\langle a}H_{b\rangle }
{}^c &=& -\c E_{ab} \,,
\label{p5}
\end{eqnarray}
and the constraint equations
\begin{eqnarray}
{\cal C}^{\bf 1}{}_a &\equiv &
\D^b\sigma_{ab} - {\ts{2\over3}}\D_a \Theta =0 \,,
\label{c1}\\
{\cal C}^{\bf 2}{}_{ab} &\equiv &
\c \sigma_{ab}- H_{ab}=0 \,,
\label{c2}\\
{\cal C}^{\bf 3}{}_a &\equiv &
\D^b E_{ab} - {\ts{1\over3}}\D_a \rho -
\ep_{abc}\sigma^b{}_d H^{cd}=0 \,,
\label{c3}\\
{\cal C}^{\bf 4}{}_a &\equiv &
\D^b H_{ab} +\ep_{abc}\sigma^b{}_d E^{cd} =0 \,.
\label{c4}
\end{eqnarray}
The propagation equations (\ref{p1})--(\ref{p5}) determine the    
covariant variables uniquely once initial data is specified on
a surface ${\cal S}(t_0)$, defined by $t=t_0$, where $t$
is comoving proper time. 
However the constraint equations (\ref{c1})--(\ref{c4}) 
place restrictions
on the initial data, and must be satisfied on ${\cal S}(t)$ 
for all $t$.
Since we have imposed the conditions
$p=0$ (dust) and $\omega_a=0$ (irrotational),\footnote{as well as 
the implicitly imposed conditions $\pi_{ab}=0$ (no
anisotropic stress) and $q_a=0$ (no energy flux)}
there is no a priori
guarantee that the constraints will not lead to inconsistencies.
However, consistency has been shown to hold 
in the generic case \cite{m}.                                      
Lengthy tensor calculations lead to the following
evolution equations of the constraints ${\cal C}^A$ along $u^a$:
\begin{eqnarray}
\dot{{\cal C}}^{\bf 1}{}_a&=&-\Theta{\cal C}^{\bf 1}{}_a+2\ep_a{}^{bc}
\sigma_b{}^d{\cal C}^{\bf 2}{}_{cd}-{\cal C}^{\bf 3}{}_a \,,
\label{pc1}\\
\dot{{\cal C}}^{\bf 2}{}_{ab}&=&-\Theta{\cal C}^{\bf 2}{}_{ab}
-\ep^{cd}{}{}_{(a}\sigma_{b)c}{\cal C}^{\bf 1}{}_d \,,
\label{pc2}\\
\dot{{\cal C}}^{\bf 3}{}_a&=&-{\ts{4\over3}}\Theta{\cal C}^{\bf 3}{}_a
+{\ts{1\over2}}\sigma_a{}^b{\cal C}^{\bf 3}{}_b-{\ts{1\over2}}\rho
{\cal C}^{\bf 1}{}_a  \nonumber\\
&&{}+{\ts{3\over2}}E_a{}^b{\cal C}^{\bf 1}{}_b 
-\ep_a{}^{bc}E_b{}^d{\cal C}^{\bf 2}
{}_{cd}+{\ts{1\over2}}\c{\cal C}^{\bf 4}{}_a \,,
\label{pc3}\\
\dot{{\cal C}}^{\bf 4}{}_a&=&-{\ts{4\over3}}\Theta{\cal C}^{\bf 4}{}_a
+{\ts{1\over2}}\sigma_a{}^b{\cal C}^{\bf 4}{}_b
 \nonumber\\
&&{}+{\ts{3\over2}}H_a{}^b{\cal C}^{\bf 1}{}_b 
-\ep_a{}^{bc}H_b{}^d{\cal C}^{\bf 2}
{}_{cd}-{\ts{1\over2}}\c{\cal C}^{\bf 3}{}_a \,.
\label{pc4}
\end{eqnarray}
It follows that if ${\cal C}^A(t_0)=0$,
then ${\cal C}^A=0$ for all time $t$. Thus, if
no further conditions (beyond $p=0$ and $\omega_a=0$) are imposed, 
{\em the constraint equations are preserved under 
evolution in the generic case.} 
We also require that the initial constraints are                   
consistent, i.e. that ${\cal C}^A(t_0)=0$ is not over-determined.
We see this as follows \cite{m}. If we freely specify, for example,
$\sigma_{ab}(t_0)$ and $\D_a\rho(t_0)$, then ${\cal C}^{\bf 1}$
determines $\D_a\Theta(t_0)$, ${\cal C}^{\bf 2}$ determines
$H_{ab}(t_0)$, and ${\cal C}^{\bf 3}$ determines $\D^bE_{ab}(t_0)$.
It could appear that ${\cal C}^{\bf 4}$ then imposes a consistency
condition, but this is not the case, since                          
it can be shown that \cite{m}
\begin{equation}
{\cal C}^{\bf 4}{}_a={\ts{1\over2}}\c{\cal C}^{\bf 1}
{}_a-\D^b{\cal C}^{\bf 2}{}_{ab} \,.
\label{i4}\end{equation}
This means that the constraint equation ${\cal C}^{\bf 4}(t_0)=0$ 
is identically satisfied
by virtue of ${\cal C}^{\bf 1}(t_0)=0={\cal C}^{\bf 2}(t_0)$. 
Thus {\em the constraint equations are
consistent with each other, and they evolve consistently,
in the generic case.}

Note that we can also interpret equation (\ref{i4}) as the        
statement that no new vector constraint arises from the 
divergence of the tensor
constraint ${\cal C}^{\bf 2}$. In other words, the tensor
constraint is essentially `transverse traceless' in content.      

When additional covariant conditions are imposed, 
then this generally valid
consistency may be disturbed and lead to integrability conditions.
Consider
the additional conditions that the gravito-electromagnetic 
tensors are divergence-free:
\[
\D^b H_{ab}=0\neq H_{ab}~\mbox{and}~\D^bE_{ab}=0\neq E_{ab} \,.
\]
These are covariant
necessary conditions for gravitational radiation in
linearized theory \cite{he}, taken over into the nonlinear regime.
Bianchi dust spacetimes were
shown to include
spatially homogeneous examples of such models in \cite{m,l},
and inhomogeneous $G_2$ solutions have also been found \cite{sv}.
However, the Newtonian condition
$H_{ab}=0$ and the anti-Newtonian condition $E_{ab}=0\neq H_{ab}$ 
both affect
the propagation equations, converting one of them into a new 
constraint. This feature, which does not arise in the cases 
$\div H=0$ or $\div E=0$, leads to complicated integrability
conditions, as described in the following sections.

\section{Newtonian-like models ($H_{ab}=0$)}

When $H_{ab}=0$, the constraints ${\cal C}^A$ are modified, 
but are still consistent, since equations (\ref{pc1})--(\ref{i4})
still hold.
But there is now an additional constraint
\[
{\cal C}^{\bf 5}{}_{ab} \equiv\c E_{ab}=0 \,,
\]
arising from the generic propagation equation (\ref{p5}), that
must be satisfied, together with its evolution along $u^a$.

First we consider 
whether the divergence of the                                       
new tensor constraint ${\cal C}^{\bf 5}$ leads to an
additional vector constraint.                                       
Using the identities (\ref{a8}) and (\ref{a14}), 
and the constraint equations (\ref{c3}) and (\ref{c4}), we 
find \cite{mn}
\begin{equation}
\D^b{\cal C}^{\bf 5}{}_{ab}={\ts{1\over2}}\c{\cal C}^{\bf 3}{}_a
-{\ts{1\over3}}\Theta{\cal C}^{\bf 4}{}_a-\sigma_a{}^b{\cal C}^
{\bf 4}{}_b \,.
\label{h1}\end{equation}
Thus no consistency condition arises from the divergence, and
$\div{\cal C}^{\bf 5}$ is determined by ${\cal C}^A$, where
$A={\bf 1},{\bf 2},{\bf 3}$, by virtue of (\ref{i4}). 

Now consider the evolution of ${\cal C}^{\bf 5}$.
Using the identities (\ref{a13.}) and
(\ref{a15}), and the propagation equations
(\ref{p4}) and (\ref{c2}), we get \cite{mn} 
\begin{eqnarray}
\dot{{\cal C}}^{\bf 5}{}_{ab} &=& 
-{\ts{4\over3}}\Theta{\cal C}^{\bf 5}{}_{ab}-{\ts{3\over2}}
\ep^{cd}{}{}_{(a}E_{b)c}{\cal C}^{\bf 1}{}_d \nonumber\\
&&{}-{\ts{1\over2}}\rho{\cal C}^{\bf 2}{}_{ab}-{\ts{3\over2}}
\ep^{cd}{}{}_{(a}\sigma_{b)c}{\cal C}^{\bf 3}{}_d+{\ts{3\over2}}
{\cal H}_{ab} \,,
\label{h2}\end{eqnarray}
where
\begin{eqnarray}
{\cal H}_{ab}&=&\ep_{cd(a}\left\{\D^e\left[E_{b)}{}^c
\sigma^d{}_e\right]+
2\D^c\left[\sigma_{b)e}E^{de}\right] \right.\nonumber\\
&&\left.{}+\sigma_{b)}{}^c\D^e
E^d{}_e+{\ts{1\over3}}\sigma^c{}_{|e|}\D^eE_{b)}{}^d\right\} \,.
\label{h3}
\end{eqnarray}
It follows that
{\em a necessary condition for consistent
evolution of the constraints in Newtonian-like (`silent')
universes is the covariant condition}
\begin{equation}
{\cal H}_{ab}=0 \,.
\label{h4}\end{equation}
This is the primary integrability condition for Newtonian-like
models. Its repeated covariant time derivatives must also be
satisfied. In \cite{5}, these derivatives, up to fourth order, 
are evaluated in a shear eigenframe, following the tetrad methods
developed in \cite{e3},
leading to a set of non-trivial conditions. 
In general, further derivatives produce 
further independent conditions,
forming an indefinite chain of integrability conditions.
The
conditions are identically satisfied in Szekeres and 
spatially homogeneous
silent models (including FLRW models, where $E_{ab}=0$), 
but not in general. Thus {\em relativistic Newtonian-like models
are in general inconsistent.\,} The same result was found
independently by different methods in \cite{s}, and is
supported by the further independent results of \cite{kp}.

Furthermore, equation (\ref{h3}) shows that the condition
(\ref{h4}) and its
time evolution are identically satisfied
in the case of covariant linearization (see \cite{eb})
around an FLRW background 
characterized by
\[
\D_a\rho=\D_a\Theta=0 ~\mbox{ and }~\sigma_{ab}=E_{ab}=H_{ab}=0\,.
\]
Thus
{\em relativistic Newtonian-like models are subject to a linearization 
instability}\,, in the sense that there exist consistent solutions
of the linearized theory of this type which are not the limit of any
consistent solution of the exact nonlinear theory.
As pointed out in \cite{5}, these results together 
cast serious doubt on
the validity and usefulness of pursuing exact `silent' solutions
as realistic models of the late universe or of gravitational
instability.
Realistic general relativistic models involve
a gravito-magnetic field, which is confirmed by the independent 
approach of \cite{kp}. 
However, as shown in the following section,
a {\em purely} gravito-magnetic field leads to even more severe
restrictions.

\section{Anti-Newtonian models ($E_{ab}=0\neq H_{ab}$)}

One of the nice features of Newtonian-like models, which facilitated
extensive investigation of their dynamics, is that the condition
$H_{ab}=0$ has the effect of decoupling the curls from the
propagation equations. The propagation equations reduce to a 
coupled system
of ordinary differential evolution equations, 
i.e. to equations (\ref{p1})--(\ref{p4}) (with $\c H_{ab}=0$ in the
last equation). Dynamical analysis of silent models
based only on these equations involves the implicit assumption
that the constraint equations are automatically satisfied. As 
shown in \cite{5,s} and outlined in
the previous section, this assumption is incorrect.

Interestingly, an analogous situation arises in the anti-Newtonian
case $E_{ab}=0\neq H_{ab}$ (although, to our knowledge, these models
have not previously been investigated).
Once again, as for silent Newtonian-like
models, the curls decouple from the
propagation equations, which again reduce to
a coupled system of ordinary differential evolution equations.
This feature reflects the fact that anti-Newtonian models are 
also `silent'.\footnote{Note that within the class of perfect 
fluid spacetimes, the covariant propagation equations (see 
\cite{mes}) reduce to ordinary differential evolution equations
under the more general conditions that $\dot{u}_a=0$
and $\c E_{ab}=0=\c H_{ab}$.}

The anti-Newtonian propagation equations are
given by (\ref{p1})--(\ref{p3}) and (\ref{p5}), with
$E_{ab}=0$ in (\ref{p3}) and 
$\c E_{ab}=0$ in (\ref{p5}).
The first 3 form a closed system determining the evolution of
$\rho$, $\Theta$, and $\sigma_{ab}$, with the last then 
determining the
propagation of $H_{ab}$ without affecting the evolution of the
other quantities. 
    Thus the matter propagation is completely decoupled from the
    Weyl tensor.\footnote{There is no Weyl tensor source term in the
    geodesic deviation equation, cf. \cite{ev}.}
Equation (\ref{p4})
      with the left hand side vanishing
is no longer a propagation 
equation -- it has become a new constraint, as in the
Newtonian-like models.
One has to investigate the consistency of the new constraint, and
this is done below. 

However, there is an added complication in the anti-Newtonian
case, not present in the Newtonian-like models.
The propagation equations are not independent, in the sense that
propagation equation (\ref{p5}) for $H_{ab}$ must be consistent
with the curl of propagation equation (\ref{p3}) 
for $\sigma_{ab}$, by virtue of the constraint equation (\ref{c2}).
Using identity (\ref{a15}), we can rewrite (\ref{p5}) as
\[
\c\dot{\sigma}_{ab}+{\ts{2\over3}}\Theta\c\sigma_{ab}
-\sigma_e{}^c\ep_{cd(a}\D^e\sigma_{b)}{}^d=0 \,,
\]
and we find that the curl of (\ref{p3}) becomes
\[ 
\c\dot{\sigma}_{ab}+{\ts{2\over3}}\Theta\c\sigma_{ab}
+\ep_{cd(a}\sigma_{b)}{}^d\D_e\sigma^{ce}
+\ep_{cd(a}\D^c\left[\sigma^{de}\sigma_{b)e}\right]=0\,,
\]
where we also used identity (\ref{a13.}) and the constraint
equation (\ref{c1}).
The difference between these equations leads to the condition
\[
\ep_{cd(a}\left\{\D^c\left[\sigma^{de}\sigma_{b)e}\right]
+\D^e\left[\sigma_{b)}{}^d\sigma^c{}_e\right]\right\}=0 \,.
\]
This condition turns out to be
an identity (satisfied by any tracefree symmetric tensor),
which is given in \cite{m}. Thus
we can ignore the propagation equation (\ref{p5}) for $H_{ab}$,
since it follows from the shear propagation equation (\ref{p3})
and the constraint equation (\ref{c1}), using covariant identities.

The coupled system of ordinary differential evolution equations
to be satisfied is then 
\begin{eqnarray}
\dot{\rho} &=& -\Theta\rho  \,,
\label{p1'}\\
\dot{\Theta}&=& -{\ts{1\over3}}\Theta^2 -{\ts{1\over2}}\rho
-\sigma_{ab}\sigma^{ab} \,,
\label{p2'}\\
\dot{\sigma}_{ab}&=& -{\ts{2\over3}}\Theta\sigma_{ab}
-\sigma_{c\langle a}\sigma_{b\rangle }{}^c  \,.
\label{p3'}
\end{eqnarray}
    These determine the evolution of the matter variables, which then 
    determine the evolution of the gravito-magnetic field through
  \begin{eqnarray}
  \dot{H}_{ab} = - \Theta H_{ab} + 3\sigma_{c\langle a}H_{b\rangle}{}^c 
  \,.\label{p5'}
  \end{eqnarray}

The constraint equation
(\ref{c4}) shows that anti-Newtonian models have $\div H=0$.
This is identically satisfied by virtue of equation (\ref{i4}), which
continues to hold when $E_{ab}=0$.
An additional constraint ${\cal C}^{\bf 5}$
arises from the gravito-electric
propagation equation (\ref{p4}). The system
of constraint equations is then
\begin{eqnarray}
{\cal C}^{\bf 1}{}_a &\equiv &
\D^b\sigma_{ab} - {\ts{2\over3}}\D_a \Theta =0 \,,
\label{c1'}\\
{\cal C}^{\bf 2}{}_{ab} &\equiv& \c\sigma_{ab}-H_{ab}=0 \,,
\label{c2'}\\
{\cal C}^{\bf 3}{}_a &\equiv &
- {\ts{1\over3}}\D_a \rho -
\ep_{abc}\sigma^b{}_d H^{cd}=0 \,,
\label{c3'}\\
    {\cal C}^{\bf 4}{}_a &\equiv &
    \D^b H_{ab} =0 \,,
    \label{c4'}\\
{\cal C}^{\bf 5}{}_{ab} &\equiv& \c H_{ab}
-{\ts{1\over2}}\rho\sigma_{ab} =0 \,.
\label{c5}
\end{eqnarray}
We can eliminate $H_{ab}$ via the constraint
equation (\ref{c2'}), which expresses the fact that the shear
is a covariant
gravito-magnetic potential. Constraint equation (\ref{c3'}) becomes
\begin{equation}
\ep_{abc}\sigma^b{}_d \c \sigma^{cd}=-{\ts{1\over3}}\D_a\rho \,.
\label{c6'}\end{equation}
Second-order derivatives then arise in
constraint equation (\ref{c5}), which may be rewritten as
\begin{equation}
\D^2\sigma_{ab}=\left({\ts{1\over2}}\rho-{\ts{1\over3}}\Theta^2+
\sigma_{cd}\sigma^{cd}\right)\sigma_{ab}-\Theta\sigma^c{}_{\langle a}
\sigma_{b\rangle c}+{\ts{3\over2}}\D_{\langle a}\D^c
\sigma_{b\rangle c} \,,
\label{c5'}\end{equation}
after using identity (\ref{a16}) for the curl of the curl of a tensor,
   where $\D^2 = \D^a \D_a$.
The constraint equation (\ref{c5'})
is a nonlinear generalization 
of the covariant Helmholtz equation. It may also be
deduced as a special case of the
nonlinear wave equation for the shear that is derived in \cite{m}.

The propagation equations (\ref{p1'})--(\ref{p3'}) provide a
unique solution for $\{\rho,\Theta,\sigma_{ab}\}$, given the
values of these quantities on an initial surface ${\cal S}(t_0)$.
As in the Newtonian-like case, the problem is that the initial
data is subject to a system of
constraint equations [(\ref{c1'})--(\ref{c5}), or equivalently 
(\ref{c1'}), (\ref{c6'}) and (\ref{c5'})], that is 
in general over-determined. Before showing this in detail, we
can see intuitively how it arises,                                
since the solution $\sigma_{ab}(t_0)$                             
of equation (\ref{c5'}),
even though it allows for arbitrary tensors of integration, will 
in general not satisfy the constraint equation (\ref{c6'}).  

We consider firstly 
whether the divergence of the new tensor constraint               
${\cal C}^{\bf 5}$ leads to an additional vector constraint.      
By the identity (\ref{a14}) and the constraint equations
(\ref{c1'}), (\ref{c3'}) and (\ref{c4}), 
we find
\begin{equation}
\D^b{\cal C}^{\bf 5}{}_{ab} = -{\ts{1\over2}}\rho{\cal C}^
{\bf 1}{}_a+{\ts{1\over3}}\Theta{\cal C}^{\bf 3}{}_a 
+{\ts{1\over2}}\c{\cal C}^{\bf 4}{}_a+{\ts{1\over9}}{\cal J}_a \,,
\label{e0}\end{equation}
where
\begin{equation}
{\cal J}_a=\Theta\D_a\rho-3\rho\D_a\Theta-{\ts{3\over2}}
\sigma_a{}^b\D_b\rho \,.
\label{e1}\end{equation}
Thus by equation (\ref{e0}) it follows that
{\em a necessary condition in anti-Newtonian universes for 
consistency of the constraints on an initial surface is
the covariant condition}
\begin{equation}
\rho\D_a\Theta={\ts{1\over3}}\Theta\D_a\rho-
{\ts{1\over2}}\sigma_a{}^b\D_b\rho \,.
\label{e1'}\end{equation}
This can be interpreted as 
an algebraic relation between the spatial gradients of
$\rho$ and $\Theta$, or we can use the constraint equations
(\ref{c1'})--(\ref{c3'}) to rewrite it as an algebraic condition on
the div and curl of the shear:
\[
9\sigma_{ab}\ep^{bcd}\sigma_c{}^e\c\sigma_{de}-
6\Theta\ep_{abc}\sigma^b{}_d\c\sigma^{cd}-4\rho\D^b\sigma_{ab}=0\,.
\]

Equation (\ref{e1'}) 
is a primary integrability condition, whose successive 
derivatives must also be satisfied. There is at least one special 
situation where this condition is identically satisfied.
If $\D_a\rho=0$, as for example
in spatially homogeneous models, then by the energy conservation 
equation (\ref{p1'}) and the identity (\ref{a7}), we 
find $\D_a\Theta=0$. It then follows that ${\cal J}_a$ is identically
zero. However, note that the new constraint equation (\ref{c5}) itself
is {\em not}
necessarily identically satisfied when $\D_a\rho=0$ -- only its 
divergence vanishes identically, as seen from equation (\ref{e0}).
Below we will find a further primary integrability condition
from (\ref{c5}) which is not identically satisfied when
$\D_a\rho=0$.
We have been unable to find spatially homogeneous solutions that
satisfy equation (\ref{c5}) when $H_{ab}\neq 0$, 
and the existence of such solutions remains
an open question. (Clearly FLRW solutions, 
with $H_{ab}=0=\sigma_{ab}$, satisfy equation (\ref{c5}) identically.)
Indeed, the only solution with $E_{ab}=0\neq H_{ab}$ known to
us is not an irrotational dust solution (see Appendix B).

In general, and especially in the more physically interesting cases,
$\D_a\rho$ is nonzero and the condition
(\ref{e1'}) is not trivial.
The evolution of integrability condition (\ref{e1'})
 along $u^a$ produces a further integrability condition.
Using identity (\ref{a7}) to commute time and space derivatives,
propagation
equations (\ref{p1'})--(\ref{p3'}) to eliminate time derivative
terms, and condition (\ref{e1'}) to eliminate $\D_a\Theta$,
 we get
\begin{equation}
\rho\D_a\left[(\sigma^2)_b{}^b\right]=
\left[-{\ts{1\over3}}\rho
+{\ts{5\over12}}(\sigma^2)_b{}^b\right]\D_a\rho
-{\ts{1\over3}}\Theta\sigma_{ab}\D^b\rho
-{\ts{1\over4}}(\sigma^2)_{\langle ab\rangle}
\D^b\rho\,,
\label{x1}\end{equation}
where $(\sigma^2)_{ab}=\sigma_a{}^c\sigma_{bc}$
is the contracted tensor product.
It is apparent that in general, with $\D_a\rho\neq0$,
condition (\ref{x1}) is not automatically
satisfied if (\ref{e1'}) is, i.e. the derived condition (\ref{x1})
is  
not an automatic consequence                                                  
of the primary condition (\ref{e1'}).
A further time derivative gives
\begin{eqnarray}
\rho\D_a\left[(\sigma^3)_b{}^b\right]&=&
\left[ -{\ts{1\over8}}\Theta
(\sigma^2)_b{}^b+{\ts{7\over16}}(\sigma^3)_b{}^b
\right]\D_a\rho  \nonumber\\
&&{}-{\ts{13\over 24}}(\sigma^2)_c{}^c\sigma_{ab}\D^b\rho
-{\ts{1\over8}}\Theta (\sigma^2)_{\langle ab\rangle}
\D^b\rho 
-{\ts{3\over16}}(\sigma^3)_{\langle ab\rangle}\D^b\rho \,,
\label{x2}\end{eqnarray}
where we used the conditions (\ref{e1'}) and (\ref{x1}).
Clearly the $N$-th time derivative leads to an integrability
condition of the form
\[
\rho\D_a\left[(\sigma^{N+1})_b{}^b\right]=A_{(N+1)}
\D_a\rho+A_{(N)}\sigma_{ab}\D^b\rho+\cdots +A_{(0)}
(\sigma^{N+1})_{\langle ab\rangle}\D^b\rho \,,
\]
where $A_{(M)}$ involves in general $\rho$, $\Theta$ and
$(\sigma^{I})_b{}^b$, $I=0,1,\cdots,M$. The algebraic invariants
of the shear are $(\sigma^2)_a{}^a$ and $(\sigma^3)_a{}^a$
(the maximal three further independent components of the shear
correspond to the rotational freedom in the choice of frame). 
The scalars
$(\sigma^N)_a{}^a$ are not all independent.

Thus there is an indefinite chain of derived integrability conditions
on ${\cal S}(t_0)$, all of which must be satisfied. At each level,
the condition does not follow automatically from lower-level
conditions. Since each
such equation involves only the initial 
data $\{\rho,\Theta, \sigma_{ab}\}$
on ${\cal S}(t_0)$, it is clear that in general 
the chain of conditions is over-determined and
will lead to inconsistencies, i.e.
{\em relativistic                                                  
anti-Newtonian universes are in general inconsistent.}             
We conjecture that 
the new constraint equation (\ref{c5}) and
the integrability conditions (\ref{e1'}), (\ref{x1}), \dots\, that
follow from it are only consistent if the shear, and hence
the gravito-magnetic field, vanishes. However, we have been
unable to prove this conjecture given the complicated nature
of the chain of integrability conditions.

This is similar to the situation in silent Newtonian-like models,
but in contrast
to the case of shear-free rotating
dust models. Using tetrad methods
as opposed to the covariant approach adopted here,
a condition similar to (\ref{e1'}) -- i.e. an algebraic
relation between spatial gradients -- was found 
for shear-free dust in \cite{e3},
and its successive time derivatives led to the conclusion that
$\Theta\omega_a=0$.
Recently, this result was regained via covariant methods
in \cite{s2} (and then generalized in \cite{sss} from dust
to vanishing fluid four-acceleration).
The crucial
simplifying factor in the shear-free case
is the central role of the naturally defined vector
$\omega_a$, as opposed to the irrotational case, where there is
no natural algebraically defined vector, and instead the
central role is played by the
tensor $\sigma_{ab}$. Furthermore, the
propagation equation for $\omega_a$ is {\em linear} in
$\omega_a$, in contrast to the shear propagation equation
(\ref{p3'}), which is nonlinear in $\sigma_{ab}$, and which gives
rise to the proliferation of terms involving 
the gradient of the trace of tensor products.

The conclusion arising from the spatial divergence of the
anti-Newtonian constraint equation (\ref{c5}) is that in general the
models are not consistent. The conjecture is that there are
{\em no} consistent
anti-Newtonian models. Both of these are reinforced by the existence
of a further chain of integrability conditions.
This arises from the {\em time} 
evolution of (\ref{c5}), which must also be
satisfied. Using 
the identities (\ref{a13.}) and (\ref{a15}), the propagation 
equations
(\ref{p1'}) and (\ref{p3'}), and the
constraint equations 
(\ref{c1'}) and (\ref{c2'}), we find 
that\footnote{ 
Instead of constraint equation (\ref{c2'}), we can also use
the gravito-magnetic propagation equation (\ref{p5'}).}
\begin{equation}
\dot{{\cal C}}^{\bf 5}{}_{ab} =
-{\ts{4\over3}}\Theta{\cal C}^{\bf 5}{}_{ab}-{\ts{3\over2}}
\ep^{cd}{}{}_{(a}H_{b)c}{\cal C}^{\bf 1}{}_d 
+{\cal E}_{ab} \,,
\label{e2}\end{equation}
where
\begin{eqnarray}
{\cal E}_{ab}&=& {\ts{1\over6}}\rho\Theta\sigma_{ab}+{\ts{1\over2}}
\rho\sigma_{c\langle a}\sigma_{b\rangle }{}^c+3H_{c\langle a}
H_{b\rangle }{}^c +3\c \left[\sigma^c{}_{(a}H_{b)c}\right] \nonumber\\
&&{}+{\ts{3\over2}}
\ep_{cd(a}H_{b)}{}^c\D^e\sigma^d{}_e-\sigma_e{}^c\ep_{cd(a}\D^e
H_{b)}{}^d
\,.\label{e3}
\end{eqnarray}
It follows that
{\em a necessary condition for consistent evolution
of the constraints in anti-Newtonian universes
is the covariant condition}
\begin{equation}
{\cal E}_{ab}=0 \,.
\label{e4}\end{equation}
Note that, in contrast to the previous integrability condition
(\ref{e1'}), this condition involves second derivatives of the shear,
given that $H_{ab}=\c\sigma_{ab}$.

Clearly this condition is identically satisfied in the             
degenerate case of $\sigma_{ab}=0$, but, in line with the
conjecture stated above, it is unlikely to be satisfied
for any shearing solutions because of the chain of further
conditions that are implied. Specifically,                         
the evolution of this integrability condition 
must also be satisfied. As before, the propagation equations
can be used to eliminate time derivatives and arrive at a 
chain of derived
integrability conditions intrinsic to ${\cal S}(t_0)$,
which places inconsistent restrictions on the initial
data $\{\rho,\Theta,\sigma_{ab}\}$ in addition to those 
remarked on above. Since inconsistency
of the anti-Newtonian models in general is already implied by
these previous restrictions, we will not give the very
complicated condition arising from $\dot{\cal E}_{ab}=0$
which, together with its time derivatives, strongly reinforces
the conclusion arrived at already.
 
The key role of the condition (\ref{e4}) emerges in the        
linearized case.                                               
Our conjecture that there are in fact no consistent anti-Newtonian
models is further reinforced by an
examination of the
linearized form of the integrability conditions.
Linearization about an FLRW universe
   (see \cite{eb}) of (\ref{e4})
shows that, in contrast to 
the Newtonian-like case, the linearized integrability
condition is non-trivial, so that not all linearized
anti-Newtonian solutions are
consistent. In fact none are consistent, since the linearized
integrability conditions are satisfied only 
when $H_{ab}=0$. Since already $E_{ab} = 0$, this is the FLRW case,
which is Newtonian-like.
This result can be derived as follows.
The linearization about an FLRW background
of (\ref{e4}) produces
\begin{equation}
\Theta\sigma_{ab}=0 \,,
\label{e5}\end{equation}
implying either $\Theta =0$ or $\sigma_{ab} = 0$.
The linearized form of the other primary integrability condition
(\ref{e1'}) is 
\[
3\rho\D_a\Theta-\Theta \D_a\rho=0 \,,
\]
which is automatically satisfied if $\Theta=0$, and 
also holds if $\sigma_{ab}=0$, since in that case the spacetime
is FLRW. However FLRW spacetimes are not anti-Newtonian,
and $\Theta=0$ implies via the linearization of propagation
equation (\ref{p2'}) that $\rho=0$, and we have ruled out 
this vacuum case
by our definition of dust universes. (Furthermore, the exact
nonlinear form of propagation equation (\ref{p2'}) shows that
$\Theta=0\neq\sigma_{ab}$ 
leads to the unphysical condition $\rho<0$.)

Thus {\em there are no linearized anti-Newtonian universes.}
The implication of this result is that it is difficult to see how 
any consistent exact anti-Newtonian solution can exist. Such a 
solution would need to have the property that it 
     cannot be linearized about an FLRW solution -- 
the solution could not admit small gravito-magnetic field or shear. 
    If such solutions existed, the FLRW solution would have 
to be an isolated point in the space of all irrotational dust
solutions with $E_{ab}=0$. 

\section{Concluding remarks}

In summary,
the overall conclusion following from the 
covariant analysis in sections 3 and 4 is that irrotational dust
universes in general relativity with realistic inhomogeneity must
have both gravito-electric and gravito-magnetic fields. The
Newtonian-like case $H_{ab}=0$ is too restrictive, supporting the
argument that there is not a straightforward relationship between
general relativistic and Newtonian universes, while the anti-Newtonian
case $E_{ab}=0\neq H_{ab}$ is even more restrictive.
Newtonian-like universes 
    (i.e. those with $H_{ab} = 0$)
are in general inconsistent, and subject  
to a linearization instability.
We showed that anti-Newtonian models are also in general 
inconsistent    
by virtue of the indefinite chain of integrability conditions
(\ref{e1'}), (\ref{x1}), \dots , which arise from the spatial
divergence of
the new tensor constraint, and which over-restrict the shear.
We conjectured that only the 
degenerate shear-free case, i.e. the FLRW
models, satisfy these integrability conditions. This conjecture is
strongly reinforced by the 
second integrability condition (\ref{e4}), which
arises from the time evolution of the new constraint, and whose
linearized form implies $\Theta\sigma_{ab}=0$, leading to the
non-existence of any 
   anti-Newtonian solutions linearized about FLRW models. 

These results extend our previous work on consistency within
the class of irrotational dust spacetimes \cite{m,mle,5,l},
using a streamlined version \cite{m,mes} of the covariant
$1+3$ formalism \cite{e}, and a systematic covariant approach
to analysing consistency \cite{m}. 
This approach parallels the tetrad methods developed in \cite{e3}
for analysing the consistency of shear-free dust spacetimes, and is
well adapted for investigating more general classes of spacetime.   

\ack
We thank Malcolm MacCallum, Colin McIntosh and Carlos Sopuerta
for helpful discussions and comments.

\newpage
\[ \]

\appendix

\section{Covariant Identities}

For convenience we collect here the necessary identities from
\cite{m}:\\

\begin{eqnarray}
\c\D_a f &=&0 \,,\label{a8}\\
\left(\D_a f\right)^{\rd} &=& \D_a \dot{f}-{\ts{1\over3}}\Theta \D_a f
-\sigma_a{}^b \D_b f \,,
\label{a7}\\
\left(\D_a V_b \right)^{\rd} &=& \D_a \dot{V}_b - {\ts{1\over3}}\Theta
\D_a V_b -\sigma_a{}^c \D_c V_b +H_a{}^d\ep_{dbc}V^c \,,
\label{a9}\\
\left(\D^b S_{ab}\right)^{\rd}&=&\D^b\dot{S}_{ab}-{\ts{1\over3}}\Theta
\D^b S_{ab}-\sigma^{bc}\D_c S_{ab}+\ep_{abc}H^b{}_dS^{cd} \,,
\label{a11.}\\
\c\left(fS_{ab}\right)&=&f\c S_{ab}+\ep_{cd(a}S_{b)}{}^d\D^c f \,,
\label{a13.}\\
\D^b\,\c S_{ab} &=& {\ts{1\over2}}\ep_{abc}
\D^b\left(\D_d S^{cd}\right)+\ep_{abc}S^b{}_d\left(
{\ts{1\over3}}\Theta\sigma^{cd}-E^{cd}\right) \nonumber \\
& &{}-\sigma_{ab}\ep^{bcd}\sigma_{ce}S^e{}_d \,,
\label{a14} \\
\left(\c S_{ab}\right)^{\rd} &=& \c \dot{S}_{ab}-{\ts{1\over3}}\Theta
\c S_{ab} \nonumber \\
& &{}-\sigma_e{}^c\ep_{cd(a}\D^e S_{b)}{}^d+3H_{c\langle a}
S_{b\rangle }{}^c \,,
\label{a15}\\
\c\c S_{ab} &=& -\D^2S_{ab}+{\ts{3\over2}}\D_{\langle a}\D^c
S_{b\rangle c}+\left(\rho-{\ts{1\over3}}\Theta^2\right)S_{ab} 
\nonumber\\
&&{}+3S_{c\langle a}\left[E_{b\rangle}{}^c -{\ts{1\over3}}\Theta
\sigma_{b\rangle}{}^c\right]+\sigma_{ab}\sigma^{cd}S_{cd} 
\nonumber\\
&&{}-\sigma^c{}_a\sigma_b{}^dS_{cd}+\sigma^c{}_{(a}S_{b)}{}^d
\sigma_{cd} \,,
\label{a16}
\end{eqnarray}
~\\

\noindent 
where $S_{ab}=S_{\langle ab\rangle }$ and $\D^2\equiv \D^a\D_a$ is 
the covariant Laplacian.

\newpage

\section{A purely gravito-magnetic solution}

In \cite{almp}, an exact solution of Einstein's field equations
with $E_{ab}=0\neq H_{ab}$ is found, apparently the first such
solution. The solution is given in the form (reversing the
signature to conform with our convention)
\begin{eqnarray}
\d s^2 &=& -x^{-2}\d u^2+\left(1-x^2\right)\d y^2-2\d u\,\d y
\nonumber\\
&&{}+x^{-2}\exp\left(-4x^2\right)\d x^2+x^{-2}\d v^2 \,,
\label{b1}\end{eqnarray}
where we have labelled the coordinates in the sequence 
$x^a=(u,y,x,v)$.
The four-velocity $u^a$ is proportional to $\ell^a+n^a$ (there is
a misprint in \cite{almp} that turns the $+$ into a $-$), 
where 
\[
\tilde{\ell}=2^{-1/2}\left[-x^{-1}\d u
+(1-x)\d y\right]\,,~\tilde{n}=2^{-1/2}\left[-x^{-1}\d u-(1+x)\d y
\right]\,,
\]
are Newman-Penrose null vectors (using the notation
$\tilde{v}=v_a\d x^a$). The Segre type of the nonzero
Ricci tensor is given in \cite{almp}
as $\{1~1~z~\bar{z}\}$, but no further discussion is given of
the properties of (\ref{b1}).

We find that
\[
\tilde{u}=-x^{-1}\d u-x\d y\,,
\]
which implies
\[
\tilde{u}\wedge\d\tilde{u}=2x^{-1}\d u\wedge\d x\wedge\d y\neq 0\,.
\]
It follows that the solution is rotating.
Furthermore, it is apparent from (\ref{b1}) that $\xi^a=\delta^a{}_0$
is a timelike Killing vector, so that the solution is stationary.
(It is not static, since $\tilde{\xi}\wedge\d\tilde{\xi}\neq 0$, or,
equivalently, since $\xi^a$ is parallel to $u^a$.)
Since $u^a=x\xi^a$, Killing's equation shows that
the solution is non-expanding and
non-shearing. Thus it is kinematically characterized by
\begin{equation}
\Theta=0\,,~\sigma_{ab}=0\,,~\omega_a \neq 0 \,.
\label{b2}\end{equation}
This solution therefore has 
   similar kinematic characteristics to 
the G\"odel solution, and we can think of it as a gravito-magnetic
counterpart of 
   that
solution, which is purely gravito-electric ($H_{ab}=0$) 
   because $\nabla_b\omega_a = 0$ 
\cite{e}.\footnote{Note 
that the G\"odel solution provides a 
counter-example to the notion that rotating matter always produces
a gravito-magnetic field \cite{mb}.}

However, unlike the G\"odel solution,
the magnetic solution (\ref{b1}) has no physically significant 
interpretation by virtue of its pathological Segre type
(see \cite{ksmh}, p. 72). For this
Segre type of the Ricci tensor, the weak energy
condition is violated, so that the energy density is negative.
Not only does the solution fall outside of the class of
irrotational dust models, but it also has an unphysical source.

\end{document}